\documentclass[]{spie}  %>>> use for US letter paper
\usepackage{hyperref}
\usepackage[]{times} 
\usepackage[]{graphicx}
\usepackage{textcomp}

\title{CYCLOPS2: the fibre image slicer upgrade for the UCLES high resolution spectrograph} 

\author{Anthony Horton\supit{a}, C.G.~Tinney\supit{b}, Scott Case\supit{a}, Tony Farrell\supit{a}, 
Luke~Gers\supit{a}, Damien Jones\supit{c}, Jon~Lawrence\supit{a}, Stan Miziarski\supit{a}, 
Nick Staszak \supit{a}, David Orr\supit{a}, Minh Vuong\supit{a}, Lew Waller\supit{a} and 
Ross~Zhelem\supit{a}
\skiplinehalf
\supit{a}Australian Astronomical Observatory, PO Box 296, Epping NSW 1710, Australia; \\
\supit{b}School of Physics, University of New South Wales NSW 2052, Australia; \\
\supit{c}Prime Optics, 17 Crescent Road, Eumundi QLD 4562, Australia
}

\authorinfo{Send correspondence to ajh@aao.gov.au}
\begin{document}

Anthony Horton et al.
"CYCLOPS2: the fibre image slicer upgrade for the UCLES high resolution 
spectrograph", Ground-based and Airborne Instrumentation for Astronomy IV, 
{\em Proc. SPIE}~{\bf 8446}, 84463A (2012)

Copyright 2012 Society of Photo Optical Instrumentation Engineers. One print or electronic copy may be made for personal use only. Systematic electronic or print reproduction and distribution, duplication of any material in this paper for a fee or for commercial purposes, or modification of the content of the paper are prohibited.

\url{http://dx.doi.org/10.1117/12.924945}

\pagebreak

  \maketitle 

%%%%%%%%%%%%%%%%%%%%%%%%%%%%%%%%%%%%%%%%%%%%%%%%%%%%%%%%%%%%% 
\begin{abstract}
CYCLOPS2 is an upgrade for the UCLES high resolution spectrograph on the Anglo-Australian
Telescope, scheduled for commissioning in semester 2012A. By replacing the 5 mirror Coud\'{e}
train with a Cassegrain mounted fibre-based image slicer CYCLOPS2 simultaneously provides
improved throughput, reduced aperture losses and increased spectral resolution. Sixteen
optical fibres collect light from a 5.0 arcsecond$^2$ area of sky and reformat it into the 
equivalent of a 0.6 arcsecond wide slit, delivering a spectral resolution of R$=70000$ and 
up to twice as much flux as the standard 1 arcsecond slit of the Coud\'{e} train. CYCLOPS2
also adds support for simultaneous ThAr wavelength calibration via a dedicated fibre. 
CYCLOPS2 consists of three main components, the fore-optics unit, fibre bundle and slit unit.
The fore optics unit incorporates magnification optics and a lenslet array and is designed to
mount to the CURE Cassegrain instrument interface, which provides acquisition, guiding and
calibration facilities. The fibre bundle transports the light from the Cassegrain focus to the
UCLES spectrograph at Coud\'{e} and also includes a fibre mode scrambler. The slit unit 
consists of the fibre slit and relay optics to project an image of the slit onto the entrance 
aperture of the UCLES spectrograph. CYCLOPS2 builds on experience with the first generation 
CYCLOPS fibre system, which we also describe in this paper. 
We present the science case for an image slicing fibre feed for echelle spectroscopy and 
describe the design of CYCLOPS and CYCLOPS2.
\end{abstract}

%>>>> Include a list of keywords after the abstract 


%%%%%%%%%%%%%%%%%%%%%%%%%%%%%%%%%%%%%%%%%%%%%%%%%%%%%%%%%%%%%
\section{INTRODUCTION}
\label{sec:intro}  % \label{} allows reference to this section

The University College London Echelle Spectrograph (UCLES)\cite{Diego1990}\ is the high
optical spectrograph of Anglo-Australian Telescope (AAT).  It has been one of the longest serving,
most successful and most productive of the AAT's instruments, and remains in demand due to its 
ability to contribute to a number of important scientific areas such as exoplanetary science, 
metallicities and abundances, and asteroseismology.  By increasing both the overall efficiency
and spectral resolution the CYCLOPS2 upgrade will ensure the continuing competitiveness of the 
UCLES spectrograph in these areas.

\subsection{Exoplanetary science}

The 763 exoplanets discovered to date (as of April 2012) have resulted in an enormous 
increase our knowledge about the formation and evolution of planetary systems. They have also 
necessitated significant re-evaluation of existing planet formation models. These exoplanets 
have been the largest single catalyst in an explosion of research in the new fields of 
exoplanetary science and astrobiology over the last decade.

Detection by the “Doppler wobble” technique has played a dominant role in this exoplanetary 
explosion.  Doppler searches provide accurate estimates of key exoplanet properties; mass 
(modulo inclination to the line of sight), period, orbital semi-major axis (ie. orbital size) and 
eccentricity.   Doppler searches have discovered 477 of the currently known exoplanets, and
been used to study the properties of 701 of the 763 total.  In the process they have told us that 
while exoplanets are not uncommon ($>7$\%\ of stars host giant planets within 5 AU) and that 
multiple planet systems exist around at least 1\%\ of stars, configurations like our own Solar 
System seem to be rare, with eccentric elliptical orbits (rather than near-circular orbits) 
being the norm for exoplanets.

These Doppler searches have come from several major groups, one of which is the 
Anglo-Australian Planet Search (AAPS). They have used UCLES extensively over the
last decade and are carrying out a long term survey to search for giant planets around more than
240 nearby Solar-type stars.  The AAPS has established itself in a crucial role, 
demonstrating extremely high long-term precision\cite{Tinney2001,Vogt2010}.

In addition to this long-term precision, experiments carried out in the last few years with 
dedicated programs targeting even higher precision.  These have demonstrated that for bright stars (V$<6.5$) 
that have high intrinsic stability exposures of 15--20 minutes on the AAT can deliver velocity
precisions at a level that makes the detection of terrestrial-mass planets in short-period orbits
feasible.

\subsection{Metallicities and abundancies}

The ancient star forming events that built up the galactic disk leave traces of their presence many 
Gyr later. Stellar substructure in the nearby disk provides a way to probe these ancient events. 
Although some of the dynamical information about these events has been lost, detailed chemical 
signatures over many chemical elements provide a way to associate or tag nearby stars to common 
ancient star forming events.  UCLES has previously be used for this kind of research using targets
identified by the RAVE survey, and in coming years UCLES will complement the GALAH
spectroscopic survey undertaken by HERMES multiobject high resolution spectrograph currently under
construction for the AAT\cite{Barden2010}.  The greater wavelength coverage and higher spectral 
resolution of UCLES will allow it to perform detailed follow up studies of unusual stars identified 
by HERMES.

\subsection{Asteroseismology}

The past few years have seen great progress in measuring oscillations in solar-type stars, 
thanks to the tremendous velocity precision being developed for hunting planets.  UCLES with 
its iodine cell plays a vital role in measuring stellar oscillations\cite{Bedding2002,Bedding2007}, not least because the 
longitude of the AAT makes it central to obtaining temporal coverage that is as continuous as 
possible. Multi-longitude observations reduce the 1/day aliasing that is a big problem for 
single-site observations.

Measuring stellar oscillations is an elegant physical experiment. A star is a gaseous 
sphere that oscillates in many different modes when excited. The oscillation frequencies depend 
on the sound speed inside the star, which in turn depends on properties such as density, 
temperature and composition. The Sun oscillates in many modes simultaneously and comparing the 
mode frequencies with theoretical calculations (helioseismology) has led to significant 
revisions to solar models. The recent revision of solar abundances poses a new challenge in which 
helioseismology is sure to play an important role.

Measuring oscillation frequencies in other stars (asteroseismology) allows us to probe their 
interiors in exquisite detail and study phenomena that do not occur in the Sun. We expect 
asteroseismology to produce major advances in our understanding of stellar structure and evolution, 
and of the underlying physical processes.

The difficulty in observing solar-like oscillations lies in their tiny amplitudes (less than a 
metre per second). Thanks to the Doppler precision provided by spectrographs such as UCLES, UVES 
and HARPS, the field of asteroseismology has finally become a reality.

The University of Sydney asteroseismology group have used UCLES in combination with UVES, HARPS or 
CORALIE to observe five stars (beta Hyi, alpha Cen A \&~B, nu Ind, Procyon). However, the data from 
UCLES are substantially inferior to those from HARPS and UVES because, although the precisions 
of all three systems are comparable, the seeing induced slit losses at the AAT substantially degrade 
the signal-to-noise ratio of the spectra. A significant increase in the throughput of UCLES 
(from CYCLOPS2) would be a tremendous advantage for this work.

%%%%%%%%%%%%%%%%%%%%%%%%%%%%%%%%%%%%%%%%%%%%%%%%%%%%%%%%%%%%%
\section{INSTRUMENT DESIGN} 
\label{sec:instrument}

%%-----------------------------------------------------------
\subsection{UCLES}
\label{sec:UCLES}

UCLES is cross-dispersed Echelle spectrograph located at the Coud\'{e} focus of the 3.9 m 
Anglo-Australian Telescope (AAT).  A variable width slit allows spectral resolutions of up to 
$R\approx 100000$ to be achieved.  To keep slit losses at acceptable levels a 
slit width of approximately 1 arcsecond is more typical, which results in a spectral 
resolution of $R\approx 45000$.  The instrument resides in an insulated but not actively 
stabilised spectrograph room.  Wavelength calibration is performed using a 
ThAr hollow cathode arc lamp and an iodine absorption cell is available for precision radial 
velocity measurements.

%%-----------------------------------------------------------
\subsection{CYCLOPS} 
\label{sec:CYCLOPS}

   \begin{figure}
   \begin{center}
   \includegraphics[width=0.9\textwidth]{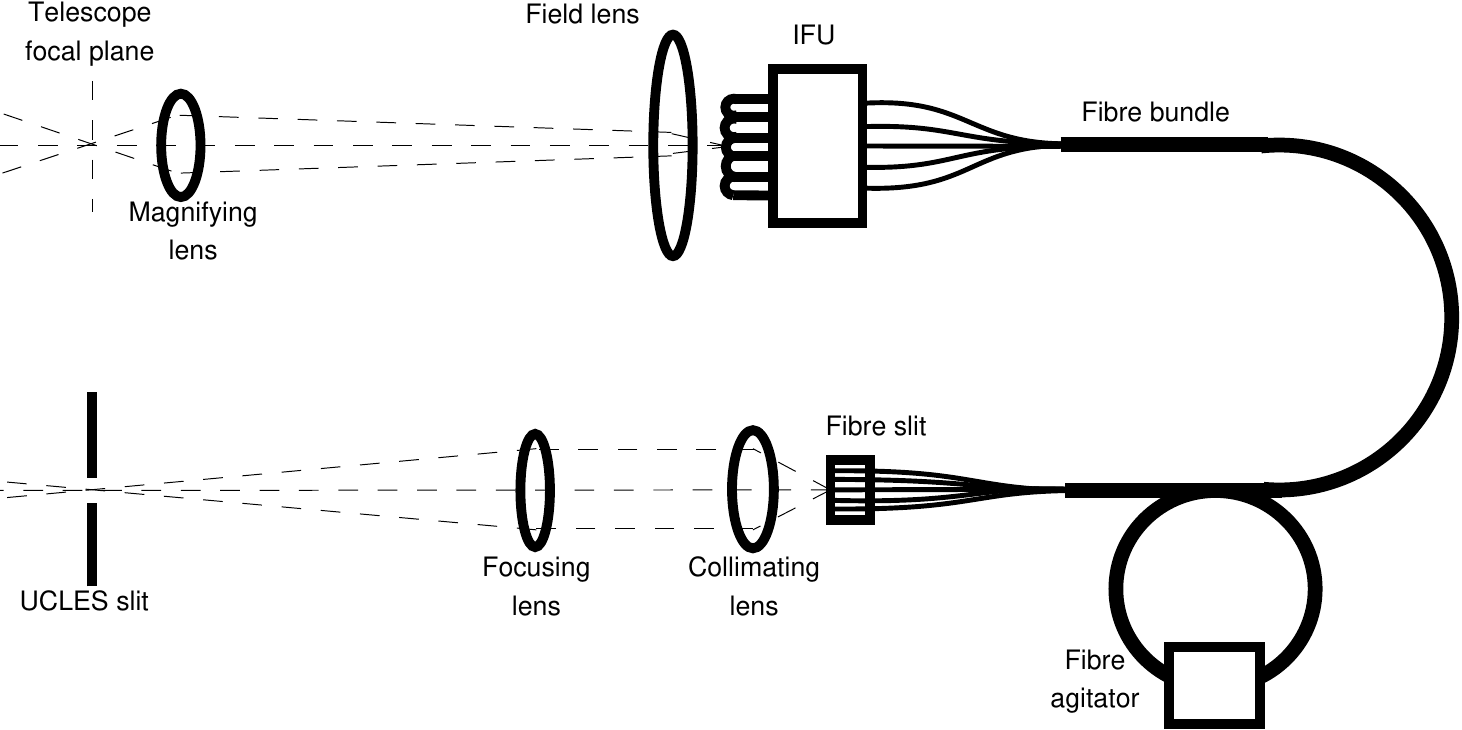}
   \end{center}
   \caption[example] 
%>>>> use \label inside caption to get Fig. number with \ref{}
   { \label{fig:CYCLOPS_schematic} 
Schematic showing the main elements of the CYCLOPS/CYCLOPS2 fibre feeds.  Not shown are the calibration, acquisition and guiding components which differ between CYCLOPS and CYCLOPS2.}
   \end{figure} 

   \begin{figure}
   \begin{center}
   \includegraphics[width=\textwidth]{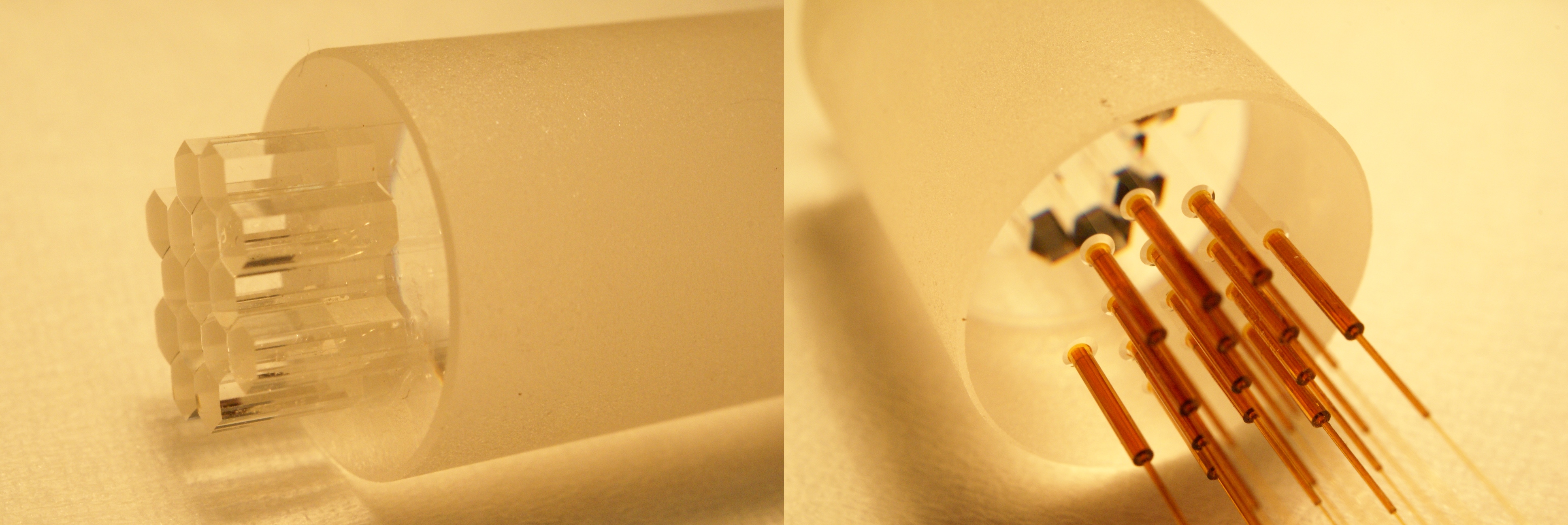}
   \end{center}
   \caption[example] 
%>>>> use \label inside caption to get Fig. number with \ref{}
   { \label{fig:CYCLOPS_IFU} 
Photographs of the CYCLOPS IFU showing the lens array attached to the front surface of the substrate (left) and the array of fibres attached to the rear surface of the substrate (right).}
   \end{figure} 

   \begin{figure}
   \begin{center}
   \includegraphics[width=0.9\textwidth]{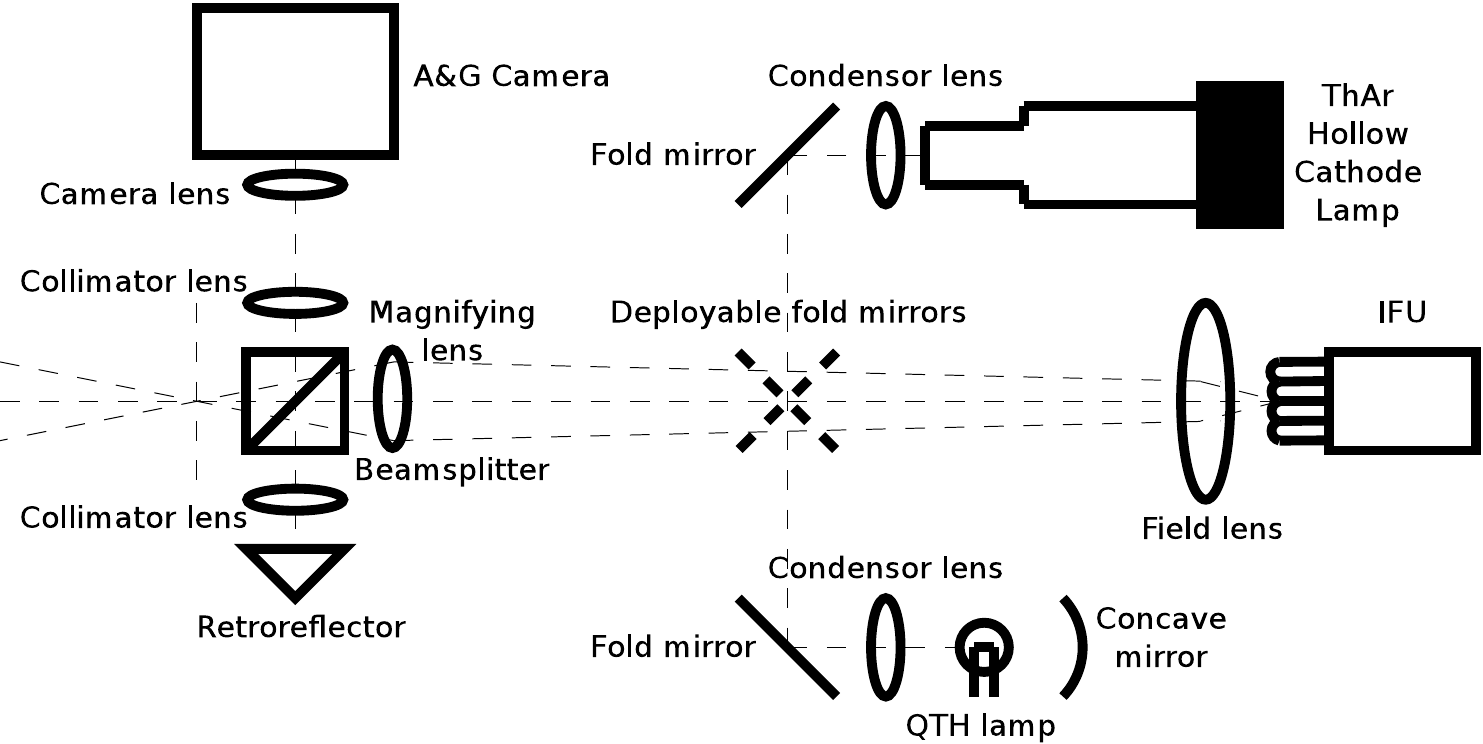}
   \end{center}
   \caption[example] 
%>>>> use \label inside caption to get Fig. number with \ref{}
   { \label{fig:CYCLOPS1_foreoptics} 
Schematic of the CYCLOPS fore optics unit including the calibration, acquisition and guiding systems.
In the CYCLOPS2 implementation these functions are provided instead by the CURE facility.}
   \end{figure}

The CYCLOPS fibre feed concept is illustrated schematically in figure \ref{fig:CYCLOPS_schematic}.  It
consists of three main components, the fore-optics unit, the fibre bundle and the slit unit.  

The fore-optics unit mounts at a Cassegrain focal station of the AAT, and
incorporates optics which project a flat, telecentric and magnified sky image onto a 
lenslet array.  The array is assembled from individual hexagonal lenslets 2 mm across (flat to 
flat) giving a spatial sampling of 0.6 arcseconds.  There are 15 lenses in total providing
a roughly circular light collecting area of 4.7 arcsecond$^2$.  The lenslet array projects an 
image on the telescope pupil onto the core of each of the 15 fibres attached to the rear of the
array substrate.  To maximise coupling efficiency each fibre was individually aligned with the
corresponding lenslet during assembly.  Figure \ref{fig:CYCLOPS_IFU} shows photographs of the
CYCLOPS integral field unit (IFU) assembly (which includes the lenslet array, fibre array and the
substrate between them) taken during integration.

The fibre bundle transports the light from the fore optics unit to the UCLES spectrograph at 
the Coud\'{e} focus position, a total distance of approximately 28 m.  The fibre used in the
current CYCLOPS implementation is Polymicro FBP custom drawn with 70/84/95 \textmu m 
core/clad/buffer diameters.  The fibres are protected by an armoured limited-bend flexible conduit.  The fibre bundle also 
incorporates a fibre agitator to prevent fibre modal noise degrading the signal-to-noise ratio of the 
spectrograph.

The slit unit terminates the fibre bundle in a close packed linear array, and includes an optical
relay that reimages this fibre pseudo-slit onto the UCLES spectrograph slit.  This relay converts
the f/5 beam exiting the fibres to an f/36 beam entering the spectrograph, resulting in an effective
slit width of 500 micron and a spectral resolution of 70000.  The length of the fibre slit is limited 
by the inter-order seperation of the UCLES spectrograph, the decision to use a close packed fibre slit
and thin fibre cladding and buffer layers were made in order to maximise the number of fibres that 
could fit into the slit.

In addition the CYCLOPS fore optics unit also contains dedicated calibration, acquisition and guiding systems.  These were included in CYCLOPS because the existing facilities available at the AAT Cassegrain focal stations did not meet the requirements of CYCLOPS.  Figure \ref{fig:CYCLOPS1_foreoptics} illustrates these components.  A beamsplitter near to the telescope focal plane redirects a few percent of the incoming light to an acquisition and guide camera.  When back-illuminated the IFU can also be seen by the acquisition and guide camera, via a reftroreflector on the opposite side of the beamsplitter.  This is used during initial setup in order to determine the approximate position of
the IFU in the camera's field of view prior to fine tuning using stellar images reconstructed from
the IFU itself.  A quartz tungsten halogen lamp for flat-fielding and a ThAr hollow cathode lamp for wavelength calibration are also included.  These illuminate the IFU using fold mirrors which can be deployed into space between the two fore optics lenses.

The CYCLOPS fibre-feed was commissioned during a series of runs in semesters 2010B and 2011B and is
now available to the community.  See \texttt{http://www.phys.unsw.edu.au/$\sim$cgt/CYCLOPS/CYCLOPS.html} for further details.

%%-----------------------------------------------------------
\subsection{CYCLOPS2 and CURE} 
\label{sec:CYCLOPS2}

  \begin{figure}
   \begin{center}
   \includegraphics[width=0.9\textwidth]{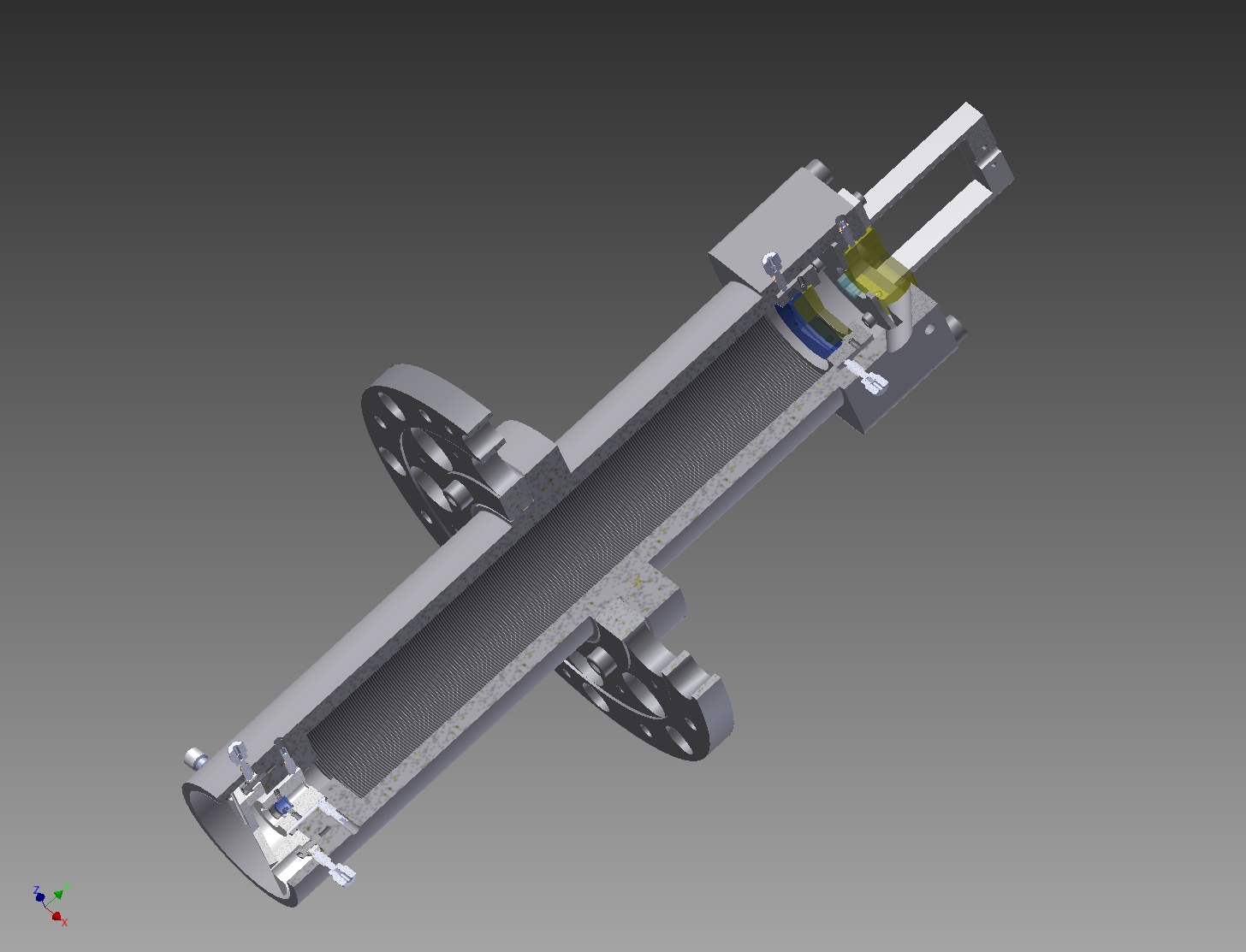}
   \end{center}
   \caption[example]
%>>>> use \label inside caption to get Fig. number with \ref{}
   { \label{fig:CYCLOPS2_foreoptics}
Mechanical model and sectioned view of the CYCLOPS2 fore-optics. From left to right optical elements include a beamsplitter (which is used to direct $\sim5$\%\ of the incoming light to the CURE acquisition and guiding camera), magnifying lens (blue), pupil stop (grey), field lens (blue and yellow), lenslet array (cyan), and substrate (yellow). The fibres are individually bonded to the rear surface of the substrate using a refined version of the techniques initially developed for CYCLOPS.  With the exception of the 
fibre agitator (not shown) CYCLOPS2 is a purely passive opto-mechanical system with no moving parts.}
   \end{figure}

   \begin{figure}
   \begin{center}
   \includegraphics[width=0.9\textwidth]{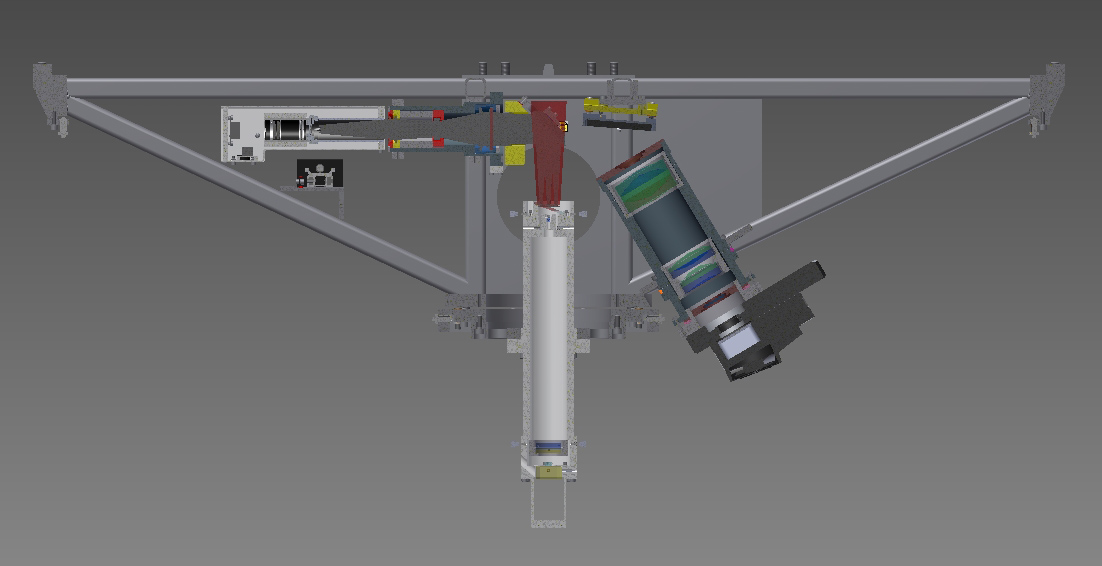}
   \end{center}
   \caption[example]
%>>>> use \label inside caption to get Fig. number with \ref{}
   { \label{fig:CURE}
Half-sectioned mechanical model of the CURE facility which mounts to the AAT's
main Cassegrain focal station.  At the bottom is the CURE instrument interface (shown with the CYCLOPS2 fore optics unit attached), on the right is the acquisition and guiding camera and on the left is the calibration assembly.}
   \end{figure}

CYCLOPS2 is a second generation fibre feed for UCLES. There 
are two main reasons for building a second fibre feed.  First, difficulties during the 
assembly of CYCLOPS resulted in 3 of the 15 fibres being damaged and as a result 
CYCLOPS has only 12 operational fibres.  This resulted in higher aperture losses than 
intended and unfortunately repair of the damaged fibres would not be possible without 
rebuilding the entire IFU-fibre bundle-fibre slit assembly.  The second reason for replacing 
CYCLOPS was in order for the fibre feed to be compatible with the CURE facility.  

CURE is a add-on to the AAT's Cassegrain instrument interface intended to simplify and 
streamline the implementation of fibre-fed and other compact instrumentation with fields of
view up to 3 arcminutes in diameter.  
It will do this by providing a suitable standardised mechanical interface as well as 
improved acquisition, guiding and calibration facilities so that CURE-compatible instruments 
will not have to incorporate their own acquistion, guiding and calibration systems as CYCLOPS
had to.  Because of this the CYCLOPS2 fore optics unit is simpler, smaller and more 
robust than the first generation CYCLOPS unit, the CYCLOPS2 design can be seen in figure \ref{fig:CYCLOPS2_foreoptics}.  The instruments expected to make use of CURE, in addition to CYCLOPS2, are 
the KOALA IFU for the AAOmega spectrograph\cite{EllisS.C.2012}, the PRAXIS OH 
suppression spectrograph\cite{Horton2012} and a number of visitor and/or experimental instruments.
Figure \ref{fig:CURE} shows the layout of the CURE facility.  The instrument interface at the base of CURE is designed to allow easy instrument exchange with repeatable alignment.  The acquisition and
guide camera looks down on the telescope focal plane via an optical relay and fold mirror, this 
'slit viewer' configuration uses either a beamsplitter 
(as in CYCLOPS2) or a mirrored entrance aperture built into the attached instrument to view the
sky.  The camera itself is a cooled interline CCD camera equipped with a set of UV, B, V, R, Ic, 
clear and blank filters, and is suitable for both rapidfire guiding exposures and deep acquisition
images.  The calibration assembly includes a quartz tungsten halogen lamp for flat fielding and
ThAr, CuAr and FeAr hollow cathode lamps for wavelength calibration.  The lamps evenly illuminate
an area of the telescope plane equivalent to 1.5 arcminutes in diameter on-sky and the relay
optics have been designed so that the illumination closely mimics that from the telescope,
including the central obstruction of the telescope pupil.  Additional calibration options are 
available using the existing discharge lamps in the AAT's acquistion and guiding unit and chimney.

\begin{figure}
   \begin{center}
   \includegraphics[angle=90,height=0.9\textheight]{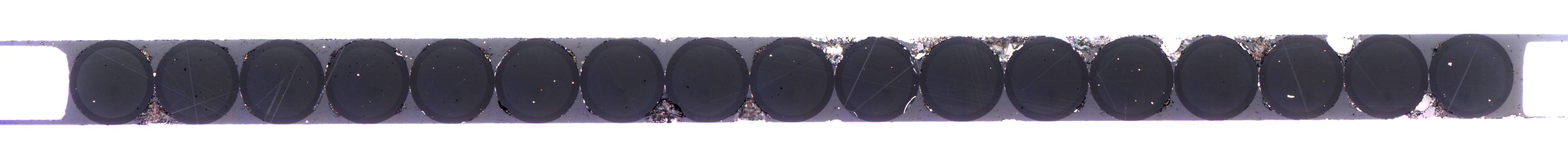}
   \end{center}
   \caption[example]
%>>>> use \label inside caption to get Fig. number with \ref{}
   { \label{fig:CYCLOPS2_slit}
Composite micrograph of the CYCLOPS2 fibre slit.}
   \end{figure}

The decision to replace CYCLOPS also provided an opportunity for upgrades compared to the 
original specifications.  It was determined that by stripping the buffer from the fibres in 
the fibre slit the fibres could be more closely packed, allowing 16 fibres to fit within 
the inter-order separation of the UCLES echellogram instead of the original 15.  It was also 
decided to include a 17\textsuperscript{th} fibre for simultaneous wavelength calibration, 
a capability that
CYCLOPS/UCLES did not have before.  Illumination of this fibre with a ThXe hollow cathode
arc lamp during science exposures will provided simultaneous wavelength calibration at the 
expense of contaminating two of the science fibres with calibration light, CYCLOPS2 will
therefore be able to operate either with 16 science fibres or with 14 science fibres plus 
simultaneous wavelength calibration.  Figure \ref{fig:CYCLOPS2_slit} is a micrograph of the CYCLOPS2
fibre slit.  The individual fibres have been close packed in direct contact with adjacent fibres and with the precision machined slit blocks on each side with the result that
the variation in fibre core positions both along and perpedicular to the slit axis are within
$\pm 1$~\textmu m.

At time of writing the CYCLOPS2 fibre-feed and the CURE facility are both in the final integration 
and assembly phase. A series of commissioning runs at the AAT are scheduled for the second half of 
2012.

 %%%%%%%%%%%%%%%%%%%%%%%%%%%%%%%%%%%%%%%%%%%%%%%%%%%%%%%%%%%%%
\section{SUMMARY} 
\label{sec:summary}

A fibre image slicer can simultaneously improve the throughput, reduce aperture losses and increase the spectral resolution of a high resolution spectrograph, significantly improving its overall performance.  The CYCLOPS2 fibre-feed, an upgrade for the UCLES high resolution spectrograph on the AAT, 
promises to be a powerful instrument for a range of science cases including Doppler exoplanet searches, stellar chemical abundance studies, and asteroseismology

%%%%%%%%%%%%%%%%%%%%%%%%%%%%%%%%%%%%%%%%%%%%%%%%%%%%%%%%%%%%%
%%%%% References %%%%%

%\bibliography{/home/ajh/Documents/Mendeley/library}   %>>>> bibliography data from Mendeley

\begin{thebibliography}{1}

\bibitem{Diego1990}
Diego, F. et~al, ``{Final tests and
  commissioning of the UCL Echelle Spectrograph},'' in {\em Instrumentation in astronomy VII}, {\em Proc.~SPIE}~{\bf 1235},  562--576 (1990).

\bibitem{Tinney2001}
Tinney, C.~G. et~al, ``{First Results from the
  Anglo‐Australian Planet Search: A Brown Dwarf Candidate and a 51 Peg–like
  Planet},'' {\em ApJ}~{\bf 551},  507--511 (2001).

\bibitem{Vogt2010}
Vogt, S.~S. et~al, ``{A super-Earth and two Neptunes
  orbiting the nearby Sun-like star 61 Virginis},'' {\em ApJ}~{\bf 708},  1366--1375 (2010).

\bibitem{Barden2010}
Barden, S.~C. et~al,
  ``{HERMES: revisions in the design for a high-resolution multi-element
  spectrograph for the AAT},'' in {\em Ground-based and Airborne
  Instrumentation for Astronomy III}, {\em Proc.~SPIE}~{\bf 7735}, 773509--773509--19 (2010).

\bibitem{Bedding2002}
Bedding, T. et~al ``{Detection of stellar oscillations with UCLES: the birth
  of asteroseismology},'' {\em AAO Newsletter} {\bf 99}, 12 (2002).

\bibitem{Bedding2007}
Bedding, T. et~al ``{Asteroseismology with UCLES},'' {\em AAO Newsletter}
  {\bf 111}, 10 (2007).

\bibitem{EllisS.C.2012}
Ellis, S.~C. et~al, ``{KOALA: a wide-field, 1000 element integral field unit for
  the Anglo-Australian Telescope},'' in {\em Ground-based and Airborne
  Instrumentation for Astronomy IV}, {\em Proc.~SPIE}~{\bf 8446},  84460V--84460V--8 (2012).

\bibitem{Horton2012}
Horton, A. et~al, ``{PRAXIS: a low
  background NIR spectrograph for fibre Bragg grating OH suppression},'' in
  {\em Modern Technologies in Space- and Ground-based Telescopes and
  Instrumentation II}, {\em Proc.~SPIE}~{\bf 8450},  84501V--84501V--6
  (2012).

\end{thebibliography}
%\bibliographystyle{spiebib}   %>>>> makes bibtex use spiebib.bst

\end{document}